\documentclass[manuscript]{aastex}
\usepackage{color}
\usepackage{url}
\usepackage{bm}

\slugcomment{to be submitted to ApJ}

\shorttitle{Study on Flare-Trigger Process with Hinode}
\shortauthors{Bamba et al.}

\begin{document}


\title{Study on Triggering Process of Solar Flares Based on Hinode/SOT Observations}


\author{Y. Bamba, K. Kusano\altaffilmark{1} and T. T. Yamamoto}
\affil{
Solar-Terrestrial Environment Laboratory, Nagoya University,
Furo-cho, Chikusa-ku, Nagoya, Aichi, 464-8601, Japan
}

\author{T. J. Okamoto}
\affil{
ISAS/JAXA, Sagamihara, Kanagawa 252-5210, Japan
}
\email{y-bamba@stelab.nagoya-u.ac.jp}


\altaffiltext{1}{Japan Agency for Marine-Earth Science and Technology (JAMSTEC),
Kanazawa-ku, Yokohama, Kanagawa, 2360001, Japan}


\begin{abstract}

We investigated four major solar flare events that occurred in active regions NOAA 10930 (December 13 and 14, 2006) and NOAA 11158 (February 13 and 15, 2011) by using data observed by the Solar Optical Telescope (SOT) onboard the Hinode satellite. To reveal the trigger mechanism of solar flares, we analyzed the spatio-temporal correlation between the detailed magnetic field structure and the emission image of the \ion{Ca}{2} H line at the central part of flaring regions for several hours prior to the onset of flares. We observed in all the flare events that the magnetic shear angle in the flaring regions exceeded $70^{\circ}$, as well as that characteristic magnetic disturbances developed at the centers of flaring regions in the pre-flare phase. These magnetic disturbances can be classified into two groups depending on the structure of their magnetic polarity inversion lines; the so-called ``Opposite-Polarity" and ``Reversed-Shear" magnetic field recently proposed by our group, \textbf{although the magnetic disturbance in one event of the four samples is too subtle to clearly recognize the detailed structure.} The result suggests that some major solar flares are triggered by rather small magnetic disturbances. We also show that the critical size of the flare-trigger field varies among flare events and briefly discuss how the flare-trigger process depends on the evolution of active regions.

\end{abstract}

\keywords{Sun: activity ― Sun: flares ― Sun: magnetic field ― Sun: sunspots ― Sun: flare trigger}

\section{Introduction} \label{sec:intro}

Solar flares are explosive phenomena driven by magnetic energy stored in the solar corona. Because interplanetary disturbances associated with solar flares sometimes impact terrestrial environments and infrastructure, understanding the flare-triggering conditions is important not only from a solar physics perspective but also for space weather forecasting. However, the onset mechanism of flares is not yet well understood, and the predictability of flare occurrence remains limited.

Many studies have suggested that various features of magnetic field are a major trigger of solar flares.
Large-scale sheared field, which may manifest as sigmoidal structures in the coronal magnetic field \citep{rustkumar96}, corresponds to free-energy storage by the non-potential magnetic field in the solar corona \citep{hagyard84}. Flux cancellations, as well as flux emerging on the magnetic polarity inversion lines (PILs), can also cause solar flares \citep{moore01}. Reversed magnetic shear is also proposed to be the cause of solar flare onset \citep{kusano04}. However, because different observations support different models, the underlying mechanism of flare onset remains elusive, and our ability to predict when flares will occur is substantially deficient.

\begin{figure}
\begin{center}
\includegraphics[height=13cm]{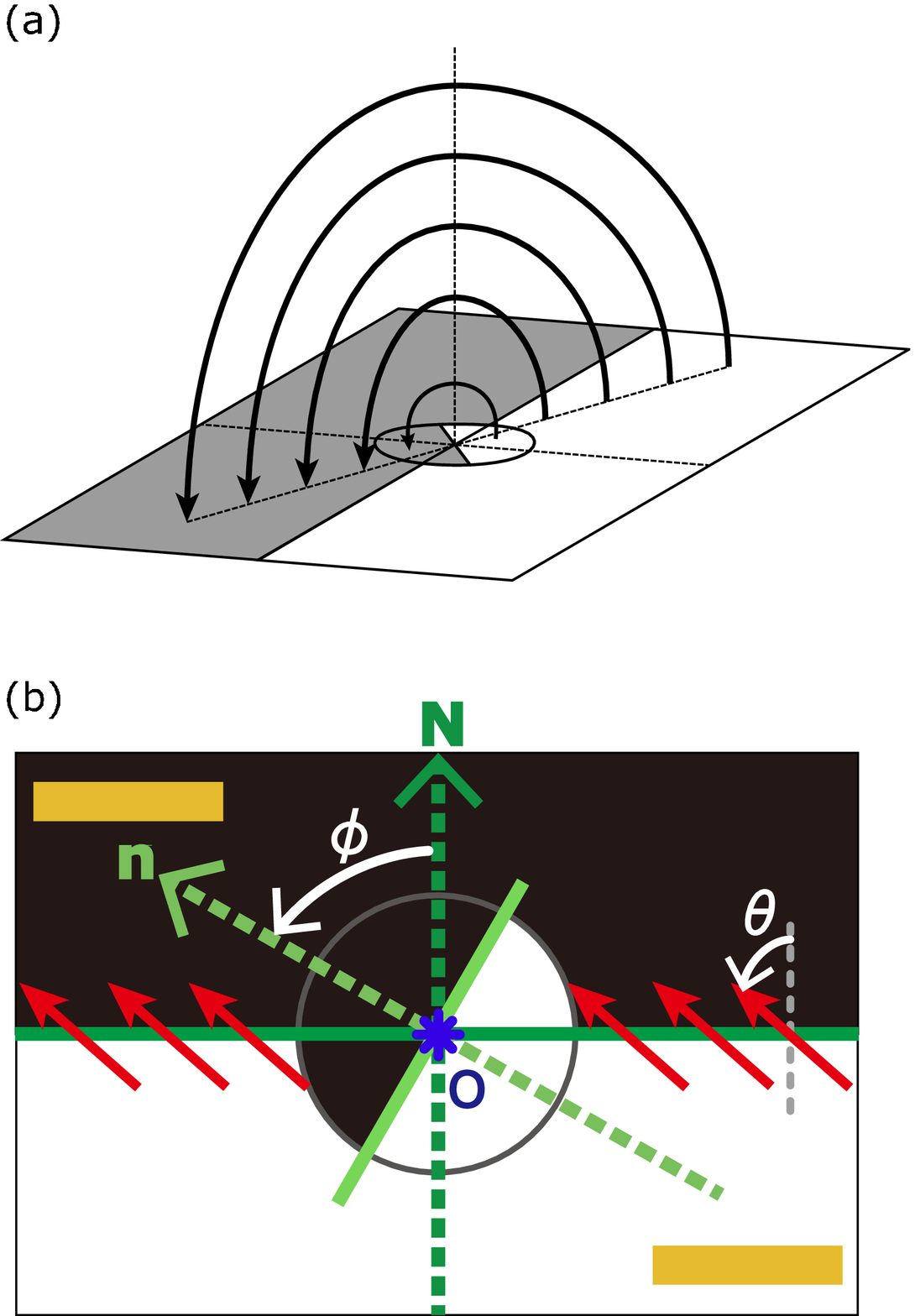}
\end{center}
\caption{
Definition of the azimuth $\phi$ and the shear angle $\theta$. 
White/black (gray) corresponds to positive/negative polarity of the LOS magnetic field. 
(a) Bird's eye view of an active region, presented in the simulation setup of \citet{kusano12}. 
(b) Top view of (a). The yellow lines indicate the sheared flare ribbon, and red arrows are the transverse magnetic field. 
We defined the center of flare-trigger region ${\it O}$ (blue asterisk), which should be located on the center of the initial flare ribbon and on the PIL. 
The vectors ${\bm N}$ and ${\bm n}$ are normal to the PIL of the averaged magnetic field and the PIL on the flare-trigger field. 
The azimuth $\phi$ is the anticlockwise angle between vectors ${\bm N}$ and ${\bm n}$. 
The shear angle $\theta$ is measured as the mean angle of the transverse magnetic field over the flare trigger region.
}
\label{fig:def_angle}
\end{figure}

Recently, \citet{kusano12} conducted ensemble simulations in which flare occurrence was systematically examined by three-dimensional magnetohydrodynamic (MHD) simulations in a wide variety of magnetic configurations. They identified two types of small magnetic disturbance, which should appear near the PIL of sheared magnetic loops as likely triggers of solar flares. These small magnetic bipole fluxes can be opposite to the major polarity (Opposite Polarity (OP) fluxes) or reversed to the averaged magnetic shear (Reversed Shear (RS) fluxes), as illustrated in Figure~\ref{fig:def_angle}(a). The simulations demonstrated that internal reconnection between these small magnetic fluxes and the sheared force-free field triggers flux rope eruption and flare reconnection, which mutually reinforce each other. Furthermore, they revealed that two major flares (the X3.4 flare on December 13, 2006 and the M6.6 flare on February 13, 2011) occurred in regions of OP- and RS-type magnetic fluxes, respectively.

The objective of this paper is to more precisely examine the model by Kusano et al. using data from four major flares observed by the Hinode satellite \citep{kosugi07}. To this end, we quantify the magnetic structures in the flaring sites. According to \citet{kusano12}, the OP or RS flare-trigger fields should exist at the center of the initial flare ribbon. Kusano and colleagues also proposed that the magnetic field in the pre-flare state can be characterized by a few parameters; the shear angle of main loop $\theta$, the azimuthal angle $\phi$, the magnetic flux of the flare-trigger field, and its displacement from the PIL.
In this study, we quantitatively analyze $\theta$ and $\phi$ of the four major flares listed in Table~\ref{table:eventlist}. All four flares occurred at regions in which the magnetic structures are consistent with the predictions of \citet{kusano12}. In addition, we estimate the critical quantity of magnetic flux required for flare triggering on the basis of magnetic field evolution data.

This paper is organized as follows. The sampled data and their analysis are presented in Sections~\ref{sec:data} and \ref{sec:method}, respectively. In Section~\ref{sec:results}, we show the results of data analyses for each of the four flare events. The relationship between the evolution of the flare-trigger field and flare onset is discussed in Section~\ref{sec:discuss}. Finally, we summarize the results in Section~\ref{sec:summary}.

\section{Data description} \label{sec:data}

Flare events were selected from the Hinode Flare Catalogue\footnote{\url{http://st4a.stelab.nagoya-u.ac.jp/hinode_flare/}} \citep{watanabe12} on the basis of the following criteria: 
\begin{enumerate}
\item The Geostationary Operational Environmental Satellite (GOES) class is larger than M5.0.
\item Events were observed by Hinode/Solar Optical Telescope \citep[SOT;][]{tsuneta08, suematsu08, ichimoto08, shimizu07} until July 31, 2011.
\item Both pre-flare and main flare phases were well covered by Hinode/SOT.
\item The flaring sites were located within $\pm75^{\circ}$ from the solar disk center.
\end{enumerate}
The above criteria were satisfied by four events occurring in active regions (ARs) NOAA 10930 and 11158; these events are analyzed in Section~\ref{sec:results}. The parameters of each event are listed in Table~\ref{table:eventlist}. Events 1 and 3 have been previously investigated by \citet{kusano12}.

AR NOAA 10930 appeared on the east limb on December 5, 2006 and produced a number of flares during disk passage. Among the four X-class flares occurring in this active region, two were observed by the Hinode/SOT, namely X3.4 at 02:14 UT December 13, 2006 \citep{kubo07} and X1.5 at 22:07 UT December 14, 2006 \citep{watanabe10}. At the time of these events, designated as Events 1 and 2, respectively, the active region was around $0^{\circ}$-$S10^{\circ}$ latitude and $W20^{\circ}$-$W45^{\circ}$ longitude.

AR NOAA 11158 appeared on the moderate southeast limb and produced one X-class flare and five M-class flares during the disc passage from February 11 to 19, 2011. We have analyzed the M6.6 flare \citep[Event 3;][]{liu12, toriumi13} and the X2.2 flare \citep[Event 4; e.g.,][]{wang12}, which were observed by the SOT at 17:28 UT February 13, 2011 and at 01:44 UT February 15, 2011, respectively. Between these two major events, numerous C-class flares occurred in this active region. During flaring, the AR was located around $S25^{\circ}$-$S15^{\circ}$ latitude and $E10^{\circ}$-$W10^{\circ}$ longitude.

Hinode/SOT is equipped with two filtergraph (FG) channels, named Broadband Filter Imager (BFI) and Narrowband Filter Imager (NFI). We use filtergrams obtained by BFI in the \ion{Ca}{2} H line (3968{\AA}) and Stokes-V/I images obtained by NFI in the Fe I line (6303{\AA}) or the Na I D1 line (5896{\AA}). \ion{Ca}{2} H line images sensitively sample the chromospheric atmosphere, whereas the Fe I and Na I D1 lines are sensitive to the line-of-sight (LOS) magnetic field in the photosphere and the base of chromosphere, respectively. The analyzed Ca-line images and Stokes-V/I images were almost continuously collected for approximately 24 hours prior to flare onset. The cadence for AR NOAA 10930 is 2 minutes, and the field of view (FOV) is 217$^{\prime\prime}$ $\times$ 108$^{\prime\prime}$ for Ca-line images and 327$^{\prime\prime}$ $\times$ 163$^{\prime\prime}$ for Stokes-V/I images. For AR NOAA 11158, the cadence is approximately 5 minutes, and FOV is 183$^{\prime\prime}$ $\times$ 108$^{\prime\prime}$ and 225$^{\prime\prime}$ $\times$ 112$^{\prime\prime}$ for Ca-line images and Stokes-V/I images, respectively.

In addition, the full polarization states (Stokes-I, Q, U, and V) of two magnetically sensitive Fe lines at 6301.5 and 6302.5 {\AA} in both active regions were observed by the Spectro-Polarimeter (SP) of SOT \citep{lites13}. In this study, the Stokes-V/I signal of filtergrams is converted to the photospheric LOS magnetic field by using the SP scan data listed in Table~\ref{table:splist}. The vector magnetogram from Hinode/SP was also used to evaluate the averaged shear angle $\theta$. The SP scan data were calibrated by the {\tt sp\_prep} procedure \citep{litesichimoto13} in the Solar Soft-Ware (SSW) package assuming a Milne-Eddington atmosphere. The inversion code MEKSY (Yokoyama et al., submitted) was adopted, and the $180^{\circ}$ ambiguity in the vector magnetograms is resolved by using the AZAM utility \citep{lites95}.

\section{Analysis methods} \label{sec:method}

\subsection{Superposition of \ion{Ca}{2} H line emission and Stakes-V/I images} \label{sec:method_1}

\citet{kusano12} proposed that internal magnetic reconnection between the flare-trigger field (which should exist in the chromosphere or in the lower corona) and the overlaying sheared magnetic field is a precursor of large flare onset. If the flare-trigger field structure is OP, the internal reconnection conjoins two sheared loops into an unstable twisted flux rope. On the other hand, if the trigger field is RS, the internal reconnection partially cancels the sheared field and causes inward collapse of the magnetic arcade as well as initiating flare reconnection. In both cases, the internal reconnection probably manifests as the pre-flare brightening of chromospheric lines. Indeed, \citet{kusano12} clearly showed that \ion{Ca}{2} H line emission in the pre-flare stage correlates with the electric current sheet formed by the sheared loop interacting with the flare-trigger field. 

Therefore, spatio-temporal correlation analysis between the magnetic field and \ion{Ca}{2} H emission is a powerful means of detecting the structure most likely to trigger a flare. To determine the flare-trigger field, we carefully superimposed Stokes-V/I images obtained by FG on Fe- or Na-lines (indicating the configuration of the photospheric LOS magnetic field) and Ca-line images. We first calibrated each image by dark-current subtraction and flat fielding by using the {\tt fg\_prep} procedure in the SSW package. We then reduced spatial fluctuations by cross-correlating two consecutive images. Selecting the Ca-line image temporally closest to the Stokes-V/I image, the two images were reconstructed to the same size because the pixel scales of BFI and NFI are different \citep{shimizu07}. Finally, we superimposed the PILs and Ca-line emission contours into the Stokes-V/I images, where the PILs are defined as the lines of zero Stokes-V/I value.

\subsection{Detection of Flare-Trigger Fields} \label{sec:method_2}

The simulation study of \citet{kusano12} predicted that flare-trigger fields of either type (OP or RS) are located between the initially brightening two-ribbons, which basically form a sheared configuration. We first examined whether a magnetic structure consistent with such a flare-trigger model exists by analyzing the superimposed Ca-line emission and Stokes-V/I images. As discussed in the next section, we identified the required flare-trigger field and the magnetic structure underlying all four events.

\subsection{Measurement of Magnetic Shear and Orientation of the Flare-Trigger Field} \label{sec:method_3}

Having determined the flare-trigger field, we measured the magnetic shear angle $\theta$ in the flaring site and the azimuth orientation $\phi$ of the flare-trigger field. \textbf{As shown in Figure~\ref{fig:def_angle}(b), $\theta$ is defined as the anticlockwise twist angle of transverse magnetic field observed by SOT/SP from the direction from the direction ${\bm N}$ normal to the averaged PIL,} defined as the line where the smoothed LOS component of the magnetic field $<B_{LOS}> = 0$. Low-pass filtering yields the smoothed magnetic field 
\begin{equation}
<B_{LOS}> = \sum_{|k_x| < k_{x0}, |k_y| < k_{y0}} \tilde{B}_{LOS}(\bm{k}) e^{i\bm{k} \cdot \bm{r}},
\label{equation:lowpass}
\end{equation}
where $\tilde{B}_{LOS}(\bm{k})$ is the complex Fourier component of mode $\bm{k}$. The critical scale $2\pi/k_{x0} = 2\pi/k_{y0} = 4.8\times10^{7}$ cm for Events 1, 2, and 4, and $2\pi/k_{x0} = 2\pi/k_{y0} = 2.4\times10^{7}$ cm for Event 3. \textbf{The shear angle $\theta$ is averaged over flaring region, as shown in the following sections.} 

\textbf{The azimuth angle of flare-trigger field $\phi$ is also defined as the anticlockwise angle between ${\bm N}$ and the local normal vector ${\bm n}$, which is orthogonal to the (non-averaged) PIL $B_{LOS}=0$. When the PIL is meandering, however, the vector ${\bm n}$ as well as angle $\phi$ is sensitive to where ${\bm n}$ is defined. In our analysis, we determined the center of flare-trigger field ${\it O}$ using Ca-line emission data.}

\textbf{According to \citet{kusano12}, solar flares can be caused by the internal magnetic reconnection between small flare-trigger field and the overlying sheared arcade, and it is likely that the internal reconnection produces the emission from chromosphere. In fact, \citet{kusano12} demonstrated that the pre-flare brightening of Ca-line can be interpreted as precursor of solar flares. Therefore, to determine the position where we should define ${\bm n}$, we sought the region of PIL where the intense emission of Ca line was observed just prior to the onset of flare. Then, assuming that the position ${\it O}$ exists within the region of PIL with Ca-line emission, we measured the average and range of ${\pm \phi}$, and adopted them as the orientation of flare-trigger field and the error range, respectively.}

\textbf{The error in $\phi$ is evaluated as the difference between minimum and maximum of multiple ${\bm n}$ measurements, whereas the error in $\theta$ is the standard deviation.}

\section{Results} \label{sec:results}

\subsection{Event 1: X3.4 flare in AR 10930 on December 13, 2006} \label{sec:results_1}

\begin{figure}
\epsscale{1.00}
\plotone{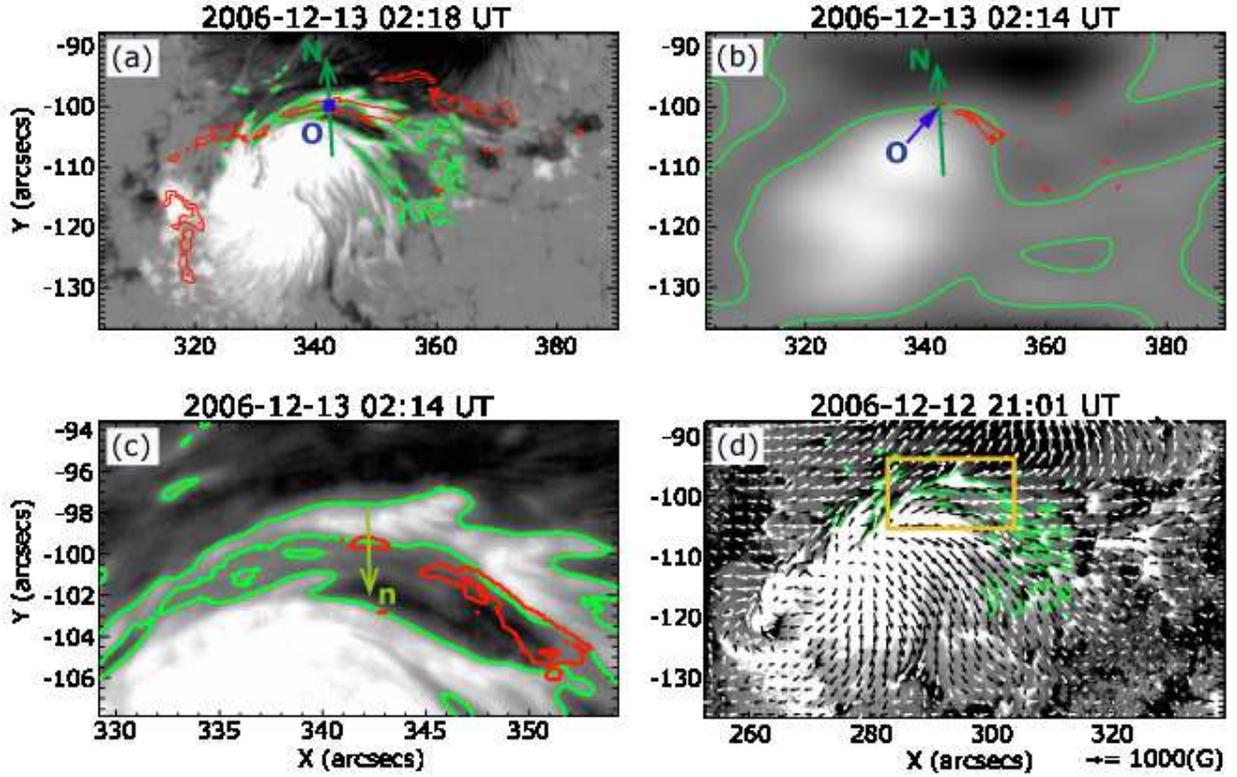}
\caption{
Images of active region NOAA 10930, and vectors ${\bm N}$ and ${\bm n}$ on the flare-trigger region in Event 1.
The grayscale part of the image corresponds to positive/negative polarity of the LOS magnetic field (Stokes-V/I), and green lines indicate the PIL. 
Red contours show the Ca-line emission. North is up and the east is to the left. 
(a) The Stokes-V/I image at 02:18 UT, when the sheared flare ribbon first appeared. 
(b) The smoothed Stokes-V/I image and the normal vector ${\bm N}$ at the point ${\it O}$. 
(c) The enlarged image of (unsmoothed) Stokes-V/I and the normal vector ${\bm n}$ at 02:14 UT. 
(d) The vector magnetic field obtained by SP at 20:30 UT December 12. 
Shear angle is calculated as the angle averaged over the yellow square.
The gray scale intensity is saturated at 0.1 (Stokes-V/I) in (a-c), and at $\pm$1000G in (d).
}
\label{fig:10930X3_four}
\end{figure}

The Stokes-V/I image corresponding to the LOS magnetic field, PIL, and Ca-line emission of the initial flare ribbon in Event 1 are plotted in Figure~\ref{fig:10930X3_four}(a). Previously, we determined that this event could be triggered by an OP-type small magnetic dipole formed between two major magnetic poles \citep[see Figure 6(b) of ][]{kusano12}. In the present study, the small-scale flare-trigger field was decomposed from the large-scale magnetic field by low-pass filtering (equation~\ref{equation:lowpass}) of the LOS magnetic field, which yielded the smoothed magnetic field $<B_{LOS}>$. In Figure~\ref{fig:10930X3_four}(b), the grayscale images corresponds to the positive/negative polarity of $<B_{LOS}>$, while green lines indicate the PILs of $<B_{LOS}>$, on which Ca-line emission (red contours) are overlaid. \textbf{The trigger point ${\it O}$ is defined as the point where the smoothed PIL is overlapped with last Ca-line emission contour at 02:14 UT December 13.} ${\it O}$ should be located at the center of the initial flare ribbon and the smoothed PIL, and the vector ${\bm N}$ is normal to the smoothed PIL at point ${\it O}$ and directed from positive to negative $<B_{LOS}>$. ${\bm N}$ denotes the reference orientation of the large-scale magnetic dipole.

\textbf{Figure~\ref{fig:10930X3_four}(c) is an enlarged view of the flare-trigger region at 02:14 UT. The vector ${\bm n}$ is defined to be normal to the (unsmoothed) PIL at last Ca-line emission contours which immediately prior to the flare onset, and directed from positive to negative $B_{LOS}$.} Figure~\ref{fig:10930X3_four}(d) shows the final vector magnetogram obtained by SOT/SP (20:30 UT on December 12) before the flaring event. The region over which the shear angle $\theta$ is averaged to derive $\theta$ is enclosed within the yellow square.

\begin{figure}
\plotone{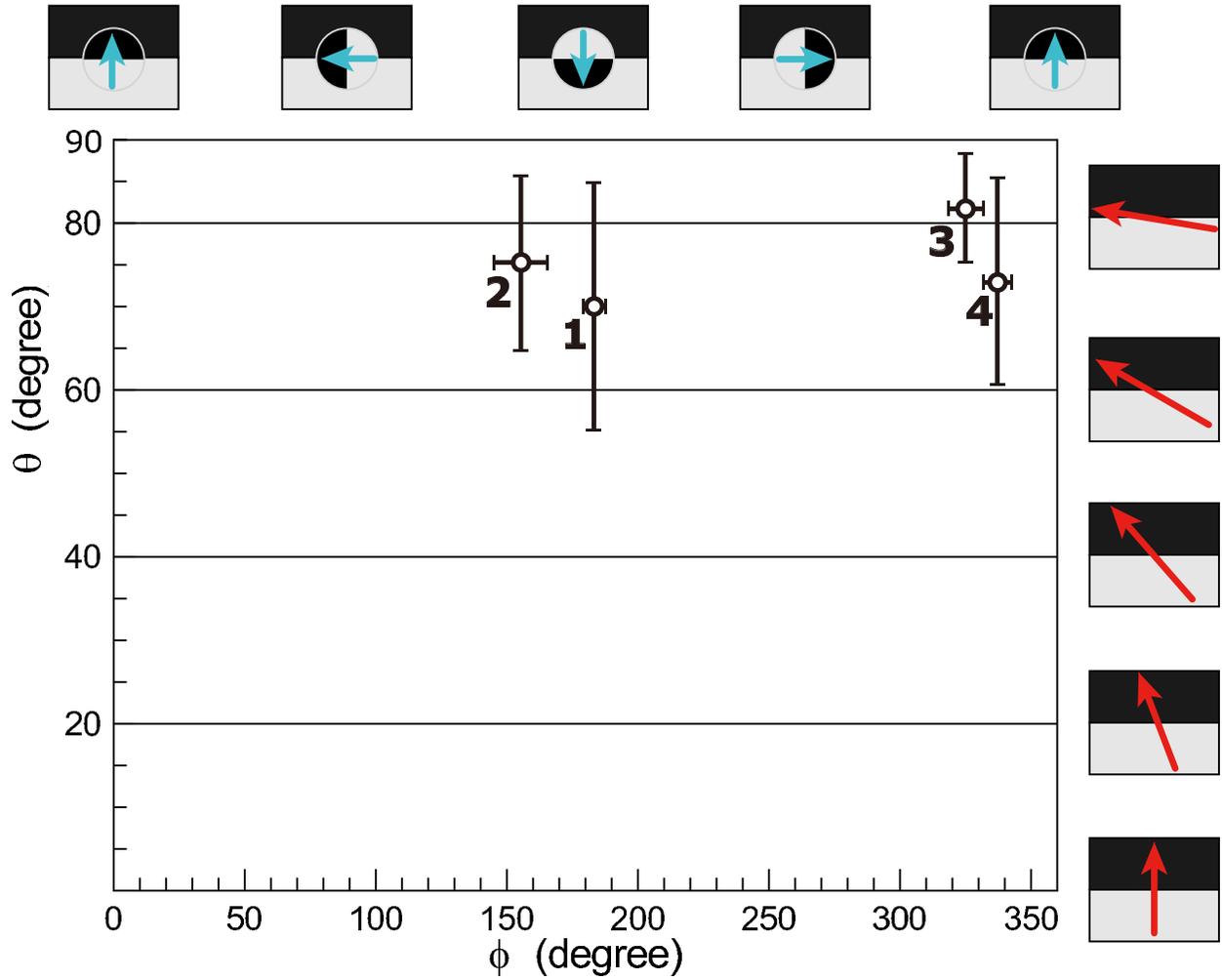}
\caption{
Summary plot of the shear angle $\theta$ and the azimuth $\phi$ for Events 1-4.
The black crosses indicate the shear angle $\theta$ and the azimuth $\phi$. 
White circles indicate the mean of $\theta$ and $\phi$ for the four cases.  
The error-bars show the standard deviation of $\theta$ and the range of $\phi$. 
Right hand and top images indicate the averaged sheared field and orientation of flare-trigger field, respectively.
}
\label{fig:diagram}
\end{figure}

The measurement results $\phi = 180^{\circ}$-$186^{\circ}$ and $\theta = 70\pm15^{\circ}$ are plotted in Figure~\ref{fig:diagram}. \textbf{Because the extent of the last Ca-line emission is different from each event, it was not able to perform statistical procedure for $\phi$.} From these results, the flare-trigger field governing Event 1 possesses an OP magnetic structure, consistent with the results of our previous paper \citep{kusano12}.

\subsection{Event 2: X1.5 flare in AR 10930 on December 14, 2006} \label{sec:results_2}

\begin{figure}
\plotone{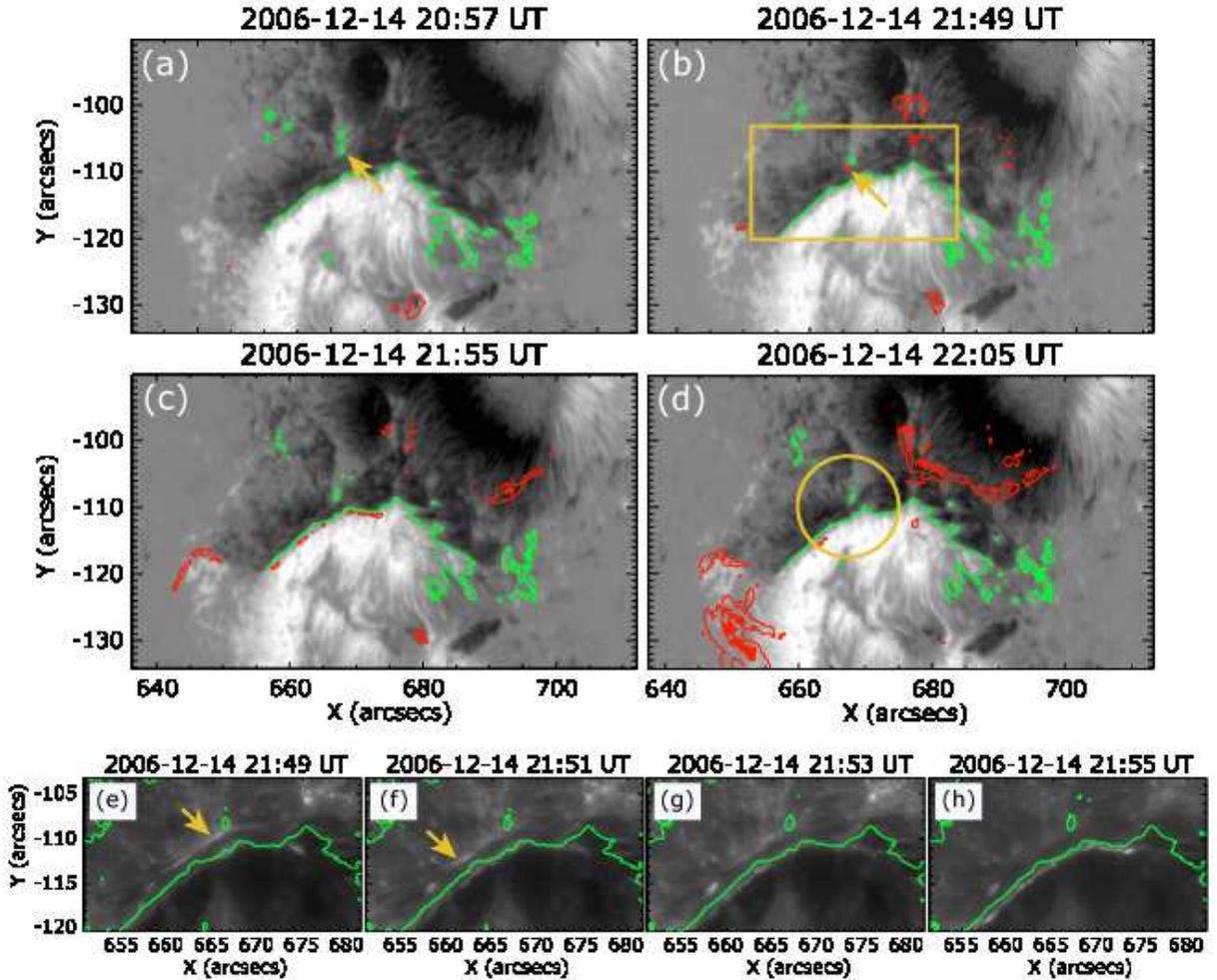}
\caption{
Temporal variations of Stokes-V/I preceding the X1.5 flare (Event 2) on December 14, 2006. 
In the top four panels, grayscale corresponds to positive/negative polarity of the LOS magnetic field, and green lines indicate the PIL.
Red contours are Ca-line emissions. 
The intensity scale saturates at $\pm0.1$ in upper four panels.
The configuration of initial flare-ribbon suggested that the trigger region is located in the yellow circle in panel (d). 
The small isolated positive island in negative sunspot and the Ca-line emission on that island are indicated by yellow arrows in panels (a) and (b), respectively. 
The bottom four panels (e-h) show the sequential images of filtergrams on the Ca-line between times for panels (b) and (c). 
The FOV corresponds to the region bordered by yellow square in panel (b), and the yellow arrows in (b) and (e) indicate the same bright point.
}
\label{fig:10930X1_dev}
\end{figure}

Figure~\ref{fig:10930X1_dev} shows the time variations of Stokes-V/I and Ca-line emission during pre-flare brightening in the X1.5 flare (Event 2) region. Green lines indicate the PILs of zero Stokes-V/I value. The structure of the initial flare ribbon is sheared (Figure~\ref{fig:10930X1_dev}(d)) and the trigger region is likely located around within the yellow circle. In fact, a small isolated positive magnetic island is detectable in the negative sunspot (indicated by the yellow arrow in Figure~\ref{fig:10930X1_dev}(a)). This island slowly approached the PIL after 20:57 UT, and a small Ca-line emission appeared on its southeast side. The small island appears to be of OP magnetic configuration. Figures~\ref{fig:10930X1_dev}(e-h) are filtergrams of the Ca-line; enlarged views of the region bordered by the yellow square in Figure~\ref{fig:10930X1_dev}(b) in which (e) and (h) were imaged at the time of (b) and (c), respectively. Yellow arrows in Figures~\ref{fig:10930X1_dev}(b) and (e) indicate the continuous bright segment, which rapidly moved southeast as shown in Figure~\ref{fig:10930X1_dev}(f). The small fiber-like structure on the Ca-line is weakened in Figure~\ref{fig:10930X1_dev}(g), whereas an elongated Ca-line brightening along the PIL emerges in Figures~\ref{fig:10930X1_dev}(h) and (c). The Ca-line emission along the PIL immediately prior to flare onset was likely caused by current sheets forming in the chromosphere as the flux rope ascended, as noted by \citet{kusano12}. From the observed progression of events (the approach of a small magnetic island, the Ca-line emissions from the island and along the PIL, and flare onset), we infer that the small magnetic island is a flare trigger.

\begin{figure}
\plotone{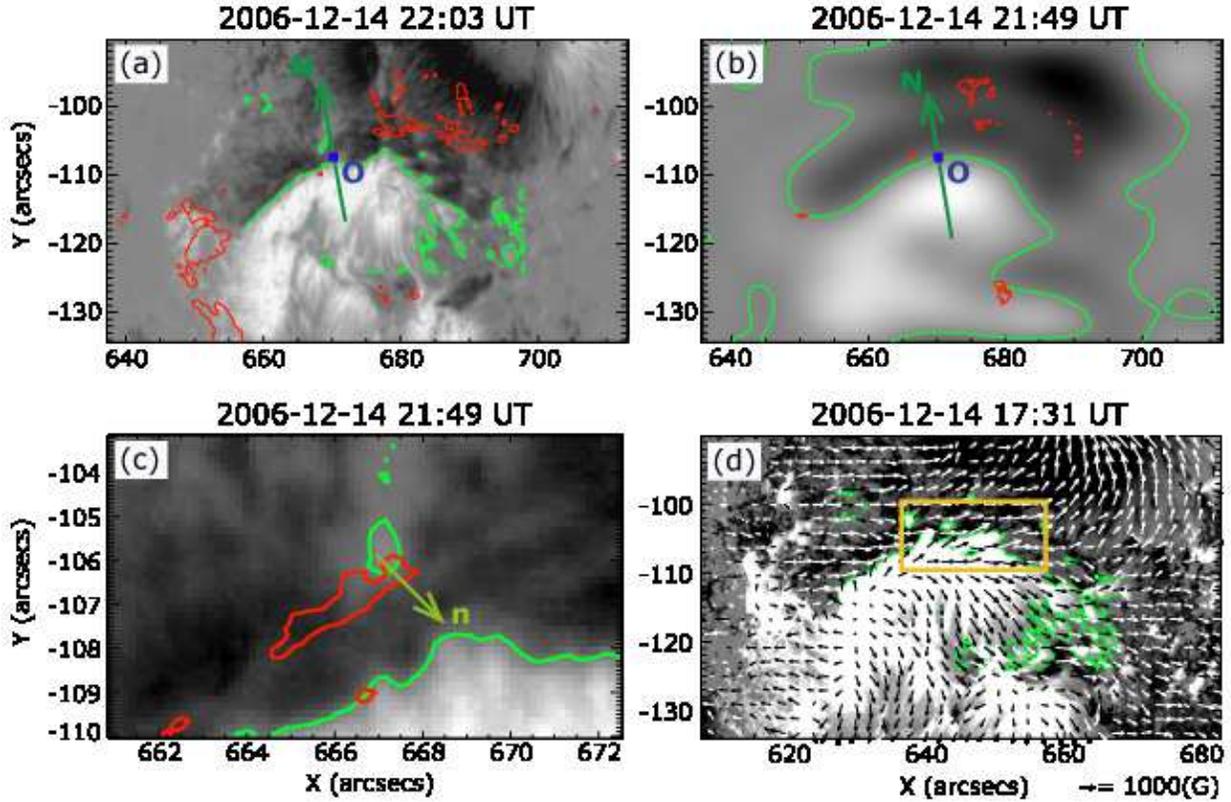}
\caption{
Images from which the azimuth $\phi$ and the shear angle $\theta$ were measured in Event 2.
The respective panels are formatted as for Figure~\ref{fig:10930X3_four}.
The trigger point ${\it O}$, vectors ${\bm N}$ and ${\bm n}$ are defined as shown in panels (a, b, and c) of Figure~\ref{fig:10930X3_four}.
Shear angle $\theta$ was measured in the region indicated by the yellow square in panel (d).
The intensity scale saturates at $\pm0.1$ in panels (a-c), and at $\pm$1000 G in panel (d).
}
\label{fig:10930X1_four}
\end{figure}

The Stokes-V/I images corresponding to the LOS magnetic field, PIL, and Ca-line emission of the initial flare ribbon in Event 2 at 22:03 UT are plotted in Figure~\ref{fig:10930X1_four}(a). Figure~\ref{fig:10930X1_four}(b) shows the smoothed Stokes-V/I image. This image is formatted identically to Figure~\ref{fig:10930X3_four}(b), and the trigger point ${\it O}$ and vector ${\bm N}$ at 21:49 UT are defined as in Figure~\ref{fig:10930X3_four}. Figure~\ref{fig:10930X1_four}(c) is an enlarged image of the trigger region, including the local normal vector ${\bm n}$. The azimuth $\phi = 145^{\circ}$-$167^{\circ}$ is the angle between vectors ${\bm N}$ and ${\bm n}$. Figure~\ref{fig:10930X1_four}(d) shows the final magnetic vector field prior to flare onset, collected by SOT/SP (17:31 UT December 14). The averaged shear angle $\theta = 75\pm11^{\circ}$.

The measured results $\phi$ and $\theta$ are plotted in Figure~\ref{fig:diagram}. \textbf{Because the magnetic island which triggered the flare is very small, the reliability of $\phi$ measurement is inferior to Event 1. However, the Ca-line emissions were bright on the near side to PIL (south side) of the small island, and the progression of the emissions shown in Figures~\ref{fig:10930X1_dev}(e-h) is consistent with the Kusano et al.’s simulation results.} Therefore, our this result suggests that the flare was triggered by the small magnetic island of OP-type configuration \citep{kusano12}.

\subsection{Event 3: M6.6 flare in AR 11158 on February 13, 2011} \label{sec:results_3}

\begin{figure}
\plotone{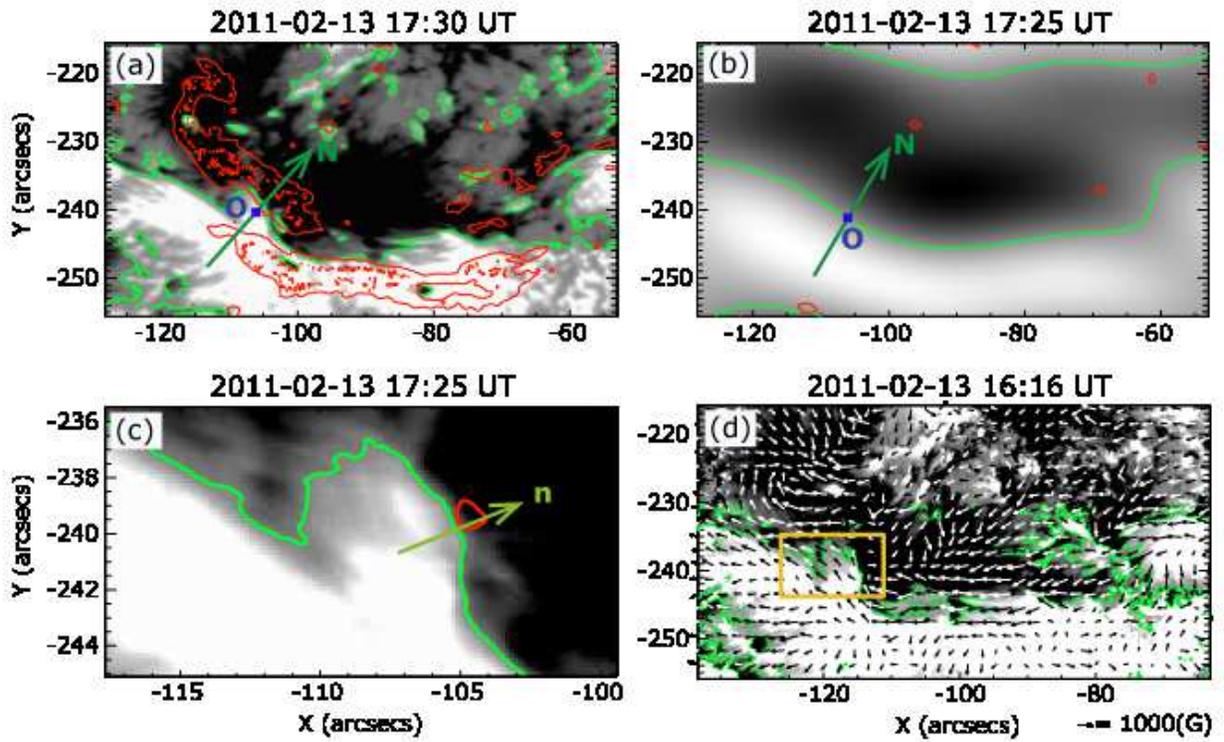}
\caption{
Images from which the azimuth $\phi$ and the share angle $\theta$ were measured in Event 3 formatted as described for Figure~\ref{fig:10930X3_four}.
The intensity scale saturates at $\pm0.1$ in panels (a-c), and $\pm$1000 G in panel (d).
}
\label{fig:11158M6_four}
\end{figure}

This flare event was also previously investigated by \citet{kusano12}, who determined that an RS magnetic configuration was the trigger for the flare \citep[see Figure 7 of][]{kusano12}. The smoothed and original Stokes-V/I images, an enlargement of the flare-trigger region, and the vector magnetogram are shown in Figure~\ref{fig:11158M6_four}. These panels in this figure are formatted identically to the corresponding panels in Figure~\ref{fig:10930X3_four}.

The origin of the flare-trigger region ${\it O}$ and normal vectors ${\bm N}$ and ${\bm n}$ at 17:25 UT are defined as shown in Figure~\ref{fig:11158M6_four}(a-c). Figure~\ref{fig:11158M6_four}(d) shows the vector magnetogram at 16:16 UT on February 13. From these images, the azimuth and shear angles were measured as $\phi = 318^{\circ}$-$331^{\circ} $ and $\theta = 82\pm7^{\circ}$. These measurements (also plotted in Figure~\ref{fig:diagram}) indicate that Event 3 was triggered by a small magnetic flux with RS characteristics, again consistent with the simulations of \citet{kusano12}.

\subsection{Event 4: X2.2 flare in AR 11158 on February 15, 2011} \label{sec:results_4}

\begin{figure}
\plotone{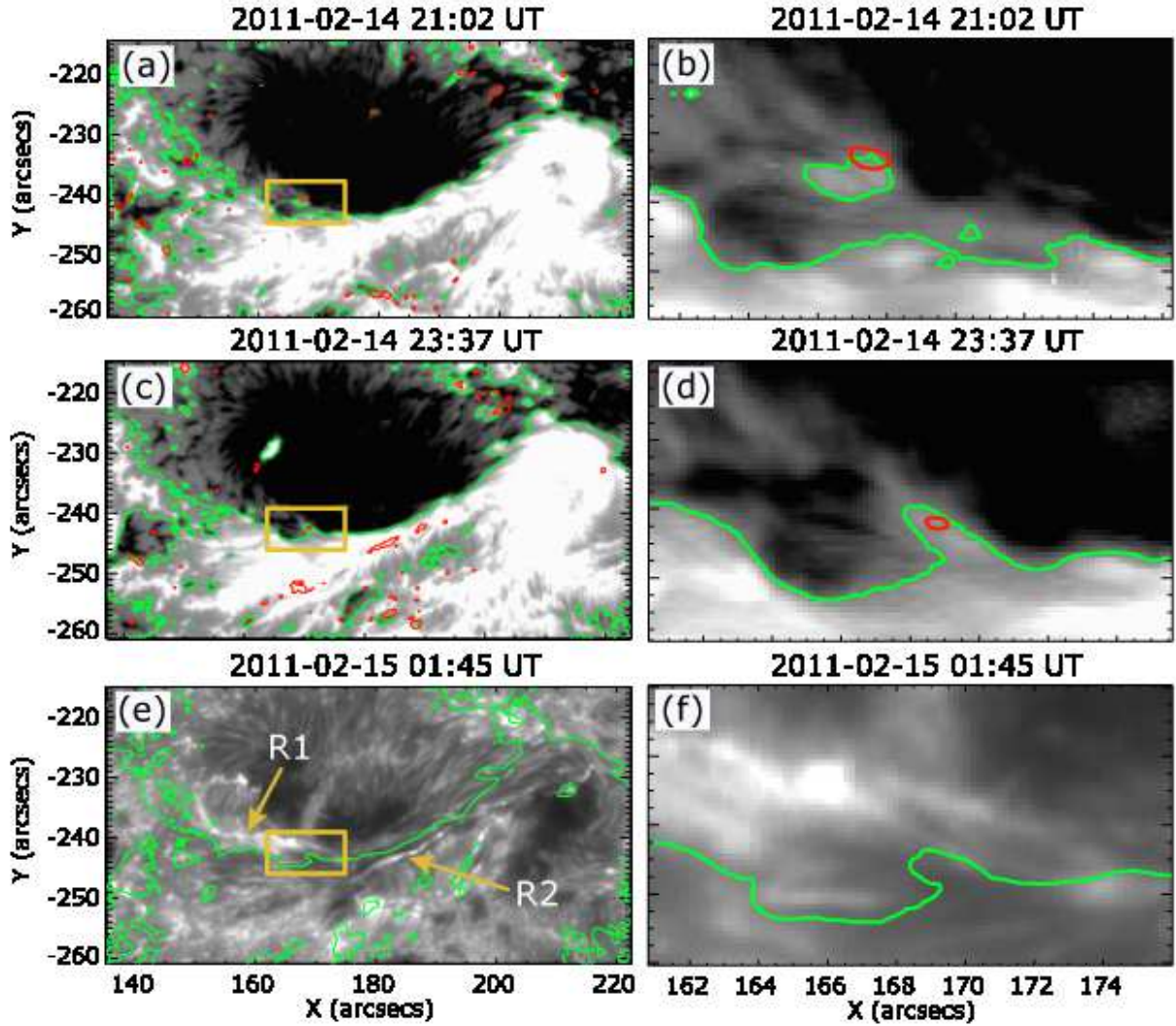}
\caption{
Time variation of Stokes-V/I preceding the X2.2 flare (Event 4) on February 15, 2011. 
Panels (a-d) and panels (e) and (f) are formatted identically to the upper- and lower four panels of Figure~\ref{fig:10930X1_dev}, respectively.
The intensity scale saturates at $\pm0.1$ in panels (a-d).
The right column shows enlarged images around the flare trigger regions (bordered by yellow squares in the left columns).
}
\label{fig:11158X2_dev}
\end{figure}

The upper four panels of Figure~\ref{fig:11158X2_dev} show the time variations of Stokes-V/I signals and Ca-line emissions in Event 4. The lower two panels show the filtergrams of the Ca-line overlaid with PILs (green lines), formatted as for the lower four panels of Figure~\ref{fig:10930X1_dev}. The right columns are enlarged images of the regions bounded by yellow squares in the left columns.

Because the initial flare ribbon has a sheared structure (R1 and R2 in Figure~\ref{fig:11158X2_dev}(e)), the trigger is inferred to locate in the region bordered by the yellow square, as shown in Figure~\ref{fig:11158X2_dev}(e). A small wedge-like structure is observed at the center of the ribbon in panel (f). This structure began as a small isolated island of positive magnetic field with Ca-line emission shown in Figures~\ref{fig:11158X2_dev}(a) and (b). This small positive island slowly developed into the small wedge-like structure of Figures~\ref{fig:11158X2_dev}(c) and (d), and the Ca-line emissions brightened on the northwest of PIL (whose polarity orientation reverses to the northeastward shear of the active region, see Figure~\ref{fig:11158X2_dev}(d)). Thus, it is assumed that this small wedge-like structure triggered the X2.2 flare.

\begin{figure}
\plotone{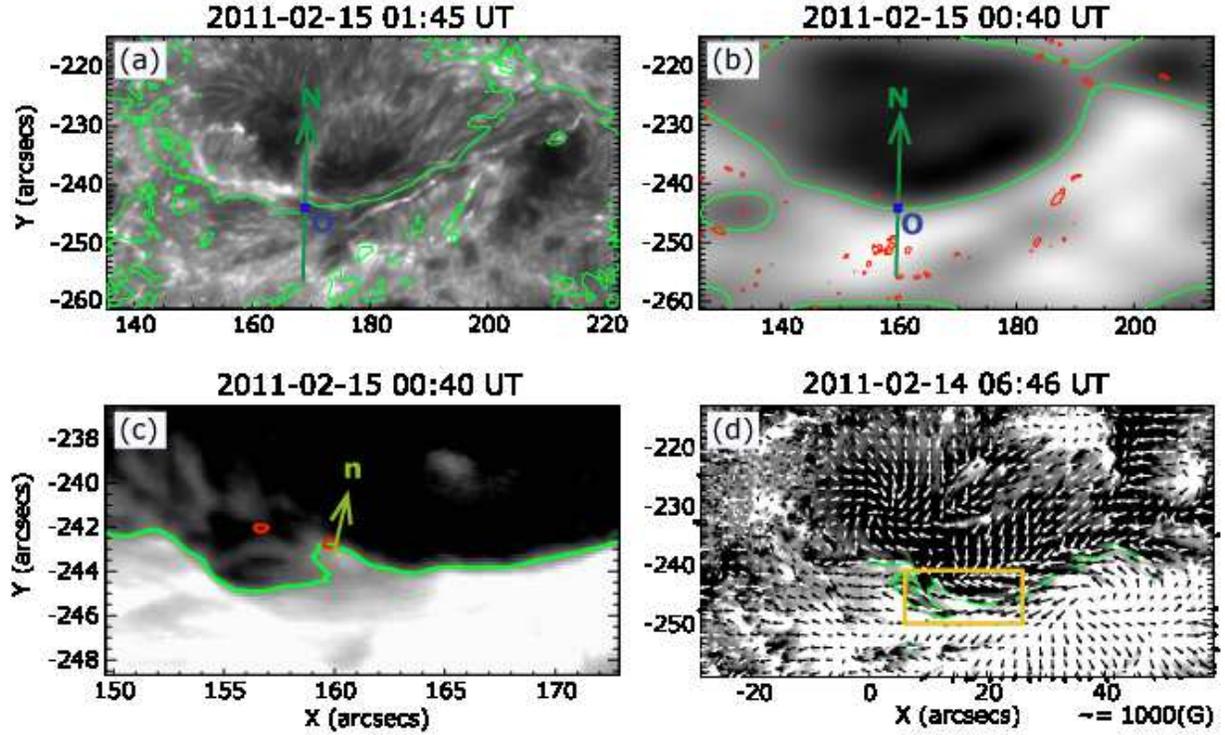}
\caption{
Images from which azimuth $\phi$ and shear angle $\theta$ were measured in Event 4.
Panels (b-d) are formatted as described for Figure~\ref{fig:10930X3_four}(b-d), respectively.
Because the initial flare ribbon was very faint, the grayscale shows  the filtergram on Ca-line in panel (a), where the green curves indicate the PILs.
The intensity scale saturates at $\pm0.1$ in panels (a-c), and $\pm$1000 G in panel (d).
}
\label{fig:11158X2_four}
\end{figure}

The trigger origin ${\it O}$ and vectors ${\bm N}$ and ${\bm n}$ at 00:40 UT are defined as in Figures~\ref{fig:11158X2_four}(b) and (c), respectively. Figure~\ref{fig:11158X2_four}(d) displays the vector magnetogram prior to flare onset, observed by SOT/SP at 06:46 UT, February 14. The yellow square delineates the region of averaged shear angle $\theta$. The azimuthal and shear angles were measured as $\phi = 334^{\circ}$-$342^{\circ} $ and $\theta = 73\pm13^{\circ}$ and are plotted in Figure~\ref{fig:diagram}. Because the SP was observed approximately one day preceding the onset of the X2.2 flare, during which time magnetic helicity was continuously injected into the active region \citep{jing12}, the measured $\theta$ likely underestimates the shear angle in the flare phase. Nevertheless, magnetic structural characteristics and the Ca-line brightening before the flare imply that the central flaring region is of RS type.

\section{Discussion} \label{sec:discuss}

\subsection{Interpretation of pre-flare emission in \ion{Ca}{2} H line} \label{sec:discuss_1}

Magnetic shear $\theta$ and the azimuthal angle of flare-trigger field $\phi$ for the four events are summarized in Figure~\ref{fig:diagram}. 
Clearly, all four events occurred in highly sheared regions ($\theta > 70^{\circ}$), consistent with previous observational studies \citep{hagyard84} and with simulations \citep{kusano12}, in which strong magnetic shear was revealed as an important condition for large flares. 
On the other hand, the four events can be classified into two groups on the basis of their azimuthal angles $\phi$. 
In Events 1 and 2, the flare-trigger field is recognizably of type OP.
Events 3 and 4 constitute another group in which the orientation of $\phi$ is weakly reversed with respect to the averaged magnetic shear. 
Although this reversed magnetic shear is rather weak compared with the flare-trigger field introduced in our simulation study \citep{kusano12}, they should nonetheless be regarded as RS triggered events, as elucidated in the following discussion.

In our analysis, the azimuth $\phi$ is defined as the angle between the normal vector ${\bm n}$ of PIL at the flare-trigger site and the orientation of the averaged magnetic polarity ${\bm N}$. 
Because the flare-trigger field is not easily deciphered from magnetic structure analysis alone, the flare-trigger site is estimated from the pre-flare Ca-line emission. 
In fact, by comparing simulations with observations, we have previously proposed that the internal reconnection that triggers flares may also induce pre-flare brightening of the Ca-line \citep{kusano12}. 
However, the simulation suggested that electric current sheets associated with internal reconnections may appear on either side as well as at the center of the flare-trigger field. 
In particular, if the flare-trigger field is small, the direction of ${\bm n}$ is highly sensitive to the origin of the flare, and the result inferred from pre-flare brightening might be misaligned. 

Despite this limitation, we emphasize that brightening immediately prior to the flares was located at one side of the peninsular structure of PIL, as seen in Figures~\ref{fig:11158M6_four}(c) and \ref{fig:11158X2_four}(c). 
In both events, the brightening appears at the side, in which the magnetic field crossing the PIL is expected to opposes the large-scale sheared magnetic field. 
Therefore, Ca-line emission and flaring are likely induced by internal reconnection between the large- and small-scale fields.
Although a more sophisticated methodology would more precisely depict the detailed structure of the flare-trigger field, our results conform to Kusano et al.'s (2012) flare-trigger scenario, in which two classes of magnetic structure can trigger flares in a highly sheared magnetic arcade.

\subsection{Critical conditions for flare triggering} \label{sec:discuss_2}

In the previous section, we demonstrated that the photospheric magnetic field of both discriminative structures (OP or RS) is indeed centralized on the flare ribbon in the four large events observed by Hinode/SOT. 
This suggests that the topological structure of the magnetic field is crucially important for flare triggering, as predicted by our previous simulation study \citep{kusano12}. 
However, the conditions under which flares erupt remain unknown, because the topological properties of magnetic field cannot explain why flares occur at specific time.

\begin{figure}
\epsscale{.70}
\plotone{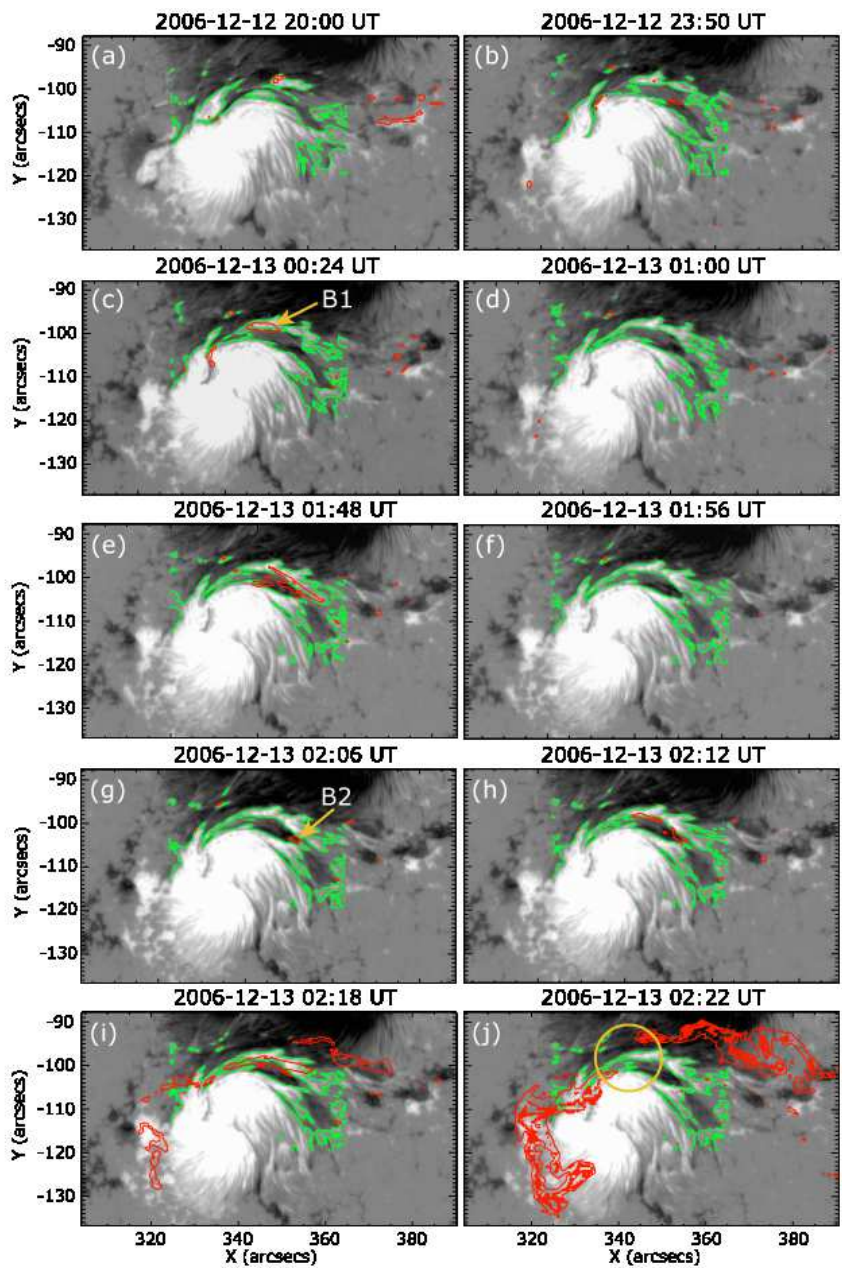}
\caption{
Temporal variations of Stokes-V/I preceding the X3.4 flare (Event 1) on December 13, 2006. 
The images are formatted as described for the top four panels of Figure~\ref{fig:10930X3_four}, and the yellow circle in panel (j) indicates the flare-trigger region. 
The transient brightening B1, located at the center of the flare-trigger region in panel (c) did not by itself induce a flare.
On the other hand, the brightening B2, located offset from the flare-trigger region, ultimately led to flare onset.
The intensity scale saturates at $\pm0.1$.
}
\label{fig:10930X3_dev}
\end{figure}

Figure~\ref{fig:10930X3_dev} displays a time series of Stokes-V/I images and \ion{Ca}{2} H emissions in AR NOAA 10930 over the four hours leading to Event 1. 
Clearly, a magnetic structure topologically consistent with the OP flare-trigger model had already appeared at 20:00 UT on December 12 (Figure~\ref{fig:10930X3_dev}(a)). 
At 00:24 UT December 13, the Ca-line emission was precisely centered on the OP-type magnetic field (indicated by B1 in Figure~\ref{fig:10930X3_dev}(c)). 
Although the geometric structures of the magnetic field and pre-flare brightening fitted the flare-trigger scenario of an OP-type magnetic field, the Ca-line emission could not be followed by flaring at this stage because it was immediately diminished (Figure~\ref{fig:10930X3_dev}(d)) and followed by another transient emission (Figure~\ref{fig:10930X3_dev}(e)). 
The other brightening occurred at 02:06 UT (B2 in Figure~\ref{fig:10930X3_dev}(g)), where it persisted and evolved to flare (Event 1). 
These results imply that the magnetic field must satisfy not only the geometrical conditions but also some additional property in order to trigger flares. 

Here, let us propose that the total magnetic flux contained in the flare-trigger field of either topology (OP or RS) contributes to the critical conditions for flare-triggering. 
This hypothesis is feasible because very small magnetic islands could not inject sufficient magnetic flux into the flux tube to destabilize it for flare onset.
To examine this hypothesis, we investigated the temporal evolution of magnetic flux in the island that formed the OP magnetic structure in Event 1. 
First, the Stokes-V/I signal was converted to magnetic intensity as explained in the Appendix.
From this conversion, we derived the LOS component of the magnetic field with a two-minute cadence and integrated it over the regions of positive magnetic polarity in the magnetic island (yellow rectangle in Figure~\ref{fig:10930_flux}(a)). 
The magnetic flux is plotted in green in Figure~\ref{fig:10930_flux}(b). 
The vertical solid line indicates the onset of flare (02:14 UT December 13), while the vertical dashed line represents the time of precedent brightening (B1 in Figure~\ref{fig:10930X3_dev}(c), 00:24 UT December 13). 
In this figure, the magnetic flux in the positive regions of the magnetic island continuously increases between 3 and 0 hours prior to flare onset. 
This conforms to our hypothesis that magnetic flux must be built to a critical level before a flare can be triggered. 

Figure~\ref{fig:10930_flux}(b) also plots the light curve of total Ca-line emission over the yellow rectangle shown in Figure~\ref{fig:10930_flux}(a). 
Here, it is noteworthy that all three pre-flare brightening events (at 00:20, 01:40, and 02:05 UT) followed a rapid magnetic flux increase denoted by yellow arrows. 
This result may be explained by assuming that rapid emerging of magnetic fluxes can favor internal reconnection with the overlying magnetic arcade. 
Therefore, the flare-trigger field must satisfy both geometrical and quantitative magnetic flux conditions.
In addition, to enable flare triggering, it should rapidly expand to drive the internal reconnection. 

Similar properties apply to Event 3.
Figure~\ref{fig:11158M6_dev} is a time series of Stokes-V/I images overlaid by Ca-line emissions in Event 3. 
Here, we observe that the growth of the wedge-like structure on the PIL (within the yellow rectangle) was followed by a flare (Figure~\ref{fig:11158M6_dev}(f)), as discussed in our previous report \citep{kusano12}. 
To quantify the growth of the flare-trigger field, we calculated the magnetic flux over the regions of positive polarity within the yellow rectangle of Figure~\ref{fig:11158M6_dev} and plotted its temporal progression.
The results, together with the light curve of Ca emission within the same area, are plotted in Figure~\ref{fig:11158_flux}, from 6 to 0 hours preceding the onset of Event 3 (marked by the vertical solid line). 
Although SOT/FG data between 13:15 and 15:00 UT are missing, the magnetic flux in the flare-trigger field clearly increased by 20-30\% between 13:15 UT and 15:20 UT. 
This process has been investigated in detail by \citet{toriumi13}.
These authors found that flux increases are associated with small positive island traveling from the north into a wedge-like region. 
The Ca-line emission on the wedge-like structure was extended northward and southward along the island displacement (indicated by the arrow in Figure~\ref{fig:11158M6_dev}(b)). 
The supply of positive flux into the wedge-like region had almost ceased before 16:30 UT, and the Ca-line emission in this region also weakened at that time. 
The Ca-line emission reappeared after 16:35 UT. 
Although the bright region on the PIL (indicated by the arrow in Figure~\ref{fig:11158M6_dev}(d)) weakly expanded until 17:05 UT, it diminished thereafter (see Figure~\ref{fig:11158M6_dev}(e)) followed by a sudden flare (Figure~\ref{fig:11158M6_dev}(f)). 

We should note that, in Event 3, magnetic flux in the flare-trigger field peaked two hours prior to flare onset (at 15:20 UT in Figure~\ref{fig:11158_flux}). 
By contrast, in Event 1, the flux in the flare-trigger field continuously increased and peaked immediately before flare onset (see Figure~\ref{fig:10930_flux}). 
This difference between Events 1 and 3 is attributable to differences in the flare-trigger processes, which are initiated by OP and RS magnetic field types, respectively. 
The former directly destabilizes the twisted flux by internal reconnection, whereas the latter indirectly forms a twisted flux by magnetic shear cancellation, and subsequent internal collapse of the magnetic arcade. 
Because reconnection between normal and reversed shear fields cannot instantaneously cancel the magnetic shear, flaring in RS fields occurs somewhat later than the evolution of magnetic flux to critical conditions. 
On the other hand, because the larger flux in OP fields can directly generate a high twisted flux, which may become destabilized by torus mode instability, flare immediately results once the flare-trigger field exceeds criticality.

However, the critical magnetic flux in the flare-trigger field differs among events. 
The flare-trigger fields in Events 2 and 4 (Figures~\ref{fig:10930X1_dev} and \ref{fig:11158X2_dev}, respectively) are much smaller than those in Events 1 and 3 (Figures~\ref{fig:10930X3_dev} and \ref{fig:11158M6_dev}, respectively), although both pairs of events occurred in the same active regions were triggered by the same type of magnetic field (OP in Events 1 and 2, RS in Events 3 and 4). 
The critical perturbation amplitude required to trigger instability depends on the proximity of the system to the unstable state. 
If the stability of the system is precarious, even small perturbations can induce instability, whereas tenaciously stable systems will not destabilize unless the perturbations are large. 

Therefore, the magnetic fields in the pre-flare phases of Events 2 and 4 were likely to be less stable than those of Events 1 and 3. 
For instance, X-ray and extreme ultraviolet observations (collected by Hinode/XRT and TRACE, respectively) revealed that a highly sheared loop was formed in the pre-flare stage of Event 2, whereas it was not seen in Event 1 \citep{su07}.
This result suggests that the sheared loop long enough to generate instability developed prior to the flare in Event 2. 

If the internal reconnection is not activated magnetic connectivity is determined by the bipolarity of the originally emerging flux, and a long sheared loop cannot fully develop. 
Eventually, a long sheared loop may be formed by the internal reconnection on the PIL. 
Therefore, the critical size of the flare-trigger field likely changes during the evolution of active regions. 
The size relationship between the flare-trigger fields of Events 1 and 2 and those of Events 3 and 4 is consistent with this notion.
However, an extended statistical study is required to elucidate the factors underlying the critical fluxes of flare-trigger fields.

\section{Summary} \label{sec:summary}

This study attempted to validate the recently proposed model of \citet{kusano12} by analyzing four major solar flares (Table~\ref{table:eventlist}) and investigating the trigger mechanisms of solar flares. 
Using the Hinode/SOT data, we derived the shear angles of large-scale magnetic fields and the azimuths of small flare-trigger fields in the four events. 
The results of this study are summarized below:
\begin{enumerate}
\item All of the flares are characterized by a ``flare-trigger field", at the center of the initial sheared flare ribbon.
\item Intermittent Ca-line brightening, which may indicate internal magnetic reconnection between the small triggering flux and the global magnetic field, was observed on the PILs of the trigger region for several hours preceding flare onset.
\item The small magnetic fluxes potentially responsible for solar flares can be classified as either ``Opposite Polarity" or ``Reversed Shear."
\item Solar flares are triggered when the small scale magnetic flux exceeds some critical level.
\item We suggest that magnetic flux must build to a critical level in the flare-trigger field before flaring can occur.
\end{enumerate}
The results of this study have enhanced our understanding of the underlying mechanisms of flare triggering, and will improve our ability to predict flare occurrence. 
However, a number of open questions remain: 
Under what conditions do flare-trigger fields develop? 
Can flare-trigger fields be incidentally or inevitably invoked? 
What determines the critical flux of a flare-trigger field? 
Which method(s) will enable reliable detection of the flare-trigger field before a flare event? 
Our study indicates that statistical analysis of different flare events based on highly accurate magnetograms and high cadence observations of chromospheric emission (such as \ion{Ca}{2} H line emissions) may be powerful enough to resolve these important questions.

\acknowledgments

Hinode is a Japanese mission developed and launched by ISAS/JAXA, which collaborates with NAOJ as a domestic partner and with NASA and STFC (UK) as international partners. 
Scientific operation of the Hinode mission is conducted by the Hinode science team organized at ISAS/JAXA. 
This team mainly consists of scientists from institutes in the partner countries. 
Support for the post-launch operation is provided by JAXA and NAOJ (Japan), STFC (UK), NASA, ESA, and NSC (Norway). 
This work was partly carried out at the NAOJ Hinode Science Center, which is supported by the Grants-in-Aid for Creative Scientific Research ``The Basic Study of Space Weather Prediction" 
(Head Investigator: K. Shibata) from the Ministry of Education, Culturem Sports, Science and Technology (MEXT), Japan, by generous donations from Sun Microsystems, and by NAOJ internal funding. 
Part of this work was carried out on the Solar Data Analysis System operated by the Astronomy Data Center in cooperation with the Hinode Science Center of the NAOJ.
This work was also supported by the Grants-in-Aid for Creative Scientific Research (B) ``Understanding and Prediction of Triggering Solar Flares" (Head Investigator: K. Kusano), from the MEXT, Japan. 
We would like to thank Shin Toriumi, Yusuke Iida, Shinsuke Imada, and Ayumi Asai for useful discussions.

\appendix

\section{Appendix} \label{sec:appendix}

\begin{figure}
\plotone{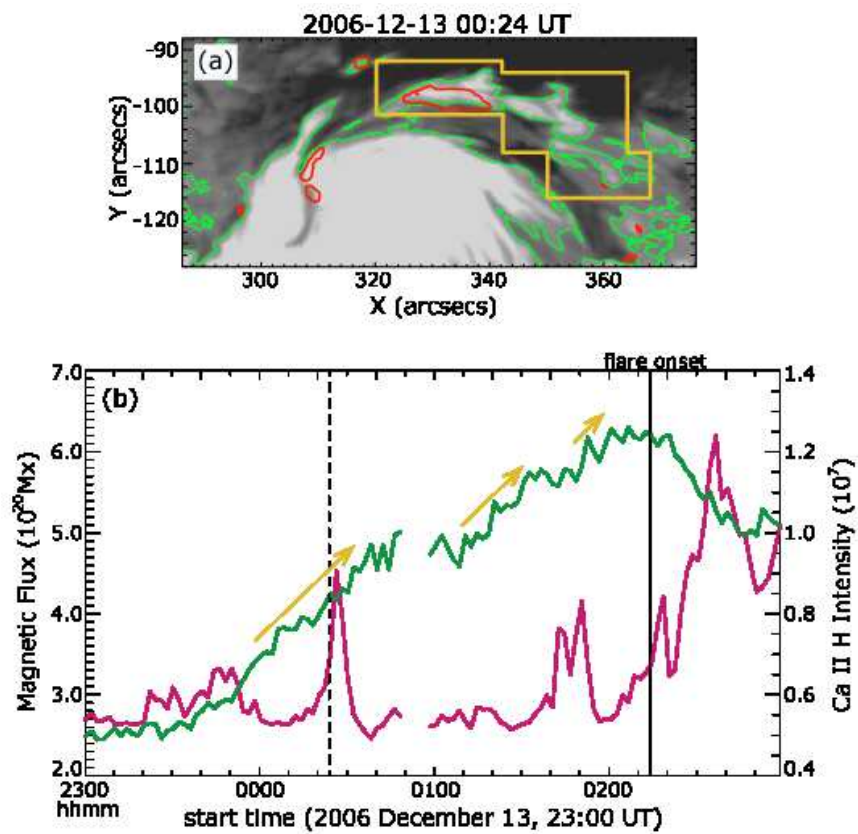}
\caption{
(a) Stokes-V/I image in the flare-trigger region of Event 1. 
Gray scale is saturated at $\pm$1000 G. 
Yellow rectangle indicates the region where magnetic flux and Ca-line intensity are integrated. 
Green lines are the PILs, and red contours indicate Ca-line emissions. 
(b) Time evolution of positive magnetic flux and Ca-line intensity integrated in the yellow rectangle in (a). 
The vertical solid line marks the onset time of the flare, 02:14 UT December 13, 2006, and the dashed line corresponds to the time of panel (a). 
Green and pink curves plot positive magnetic flux and Ca-line intensity, respectively.
}
\label{fig:10930_flux}
\end{figure}

\begin{figure}
\plotone{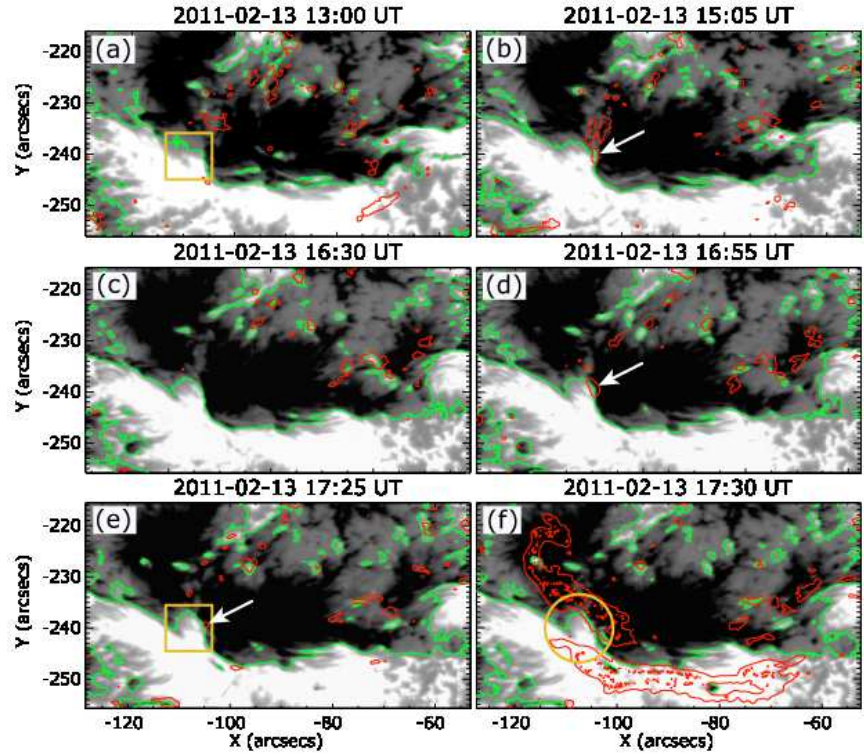}
\caption{
Temporal variations of Stokes-V/I preceding the M6.6 flare (Event 3) on February 13, 2011.	
Images are formatted as described for the top four panels of Figure~\ref{fig:10930X1_dev}, and the yellow circle in panel (f) indicates the flare-trigger region. 
The yellow squares in panels (a) and (e) identify the regions of positive magnetic flux and Ca-line intensity (plotted in Figure~\ref{fig:11158_flux}).
The gray scale intensity saturates at $\pm0.1$.
}
\label{fig:11158M6_dev}
\end{figure}

\begin{figure}
\plotone{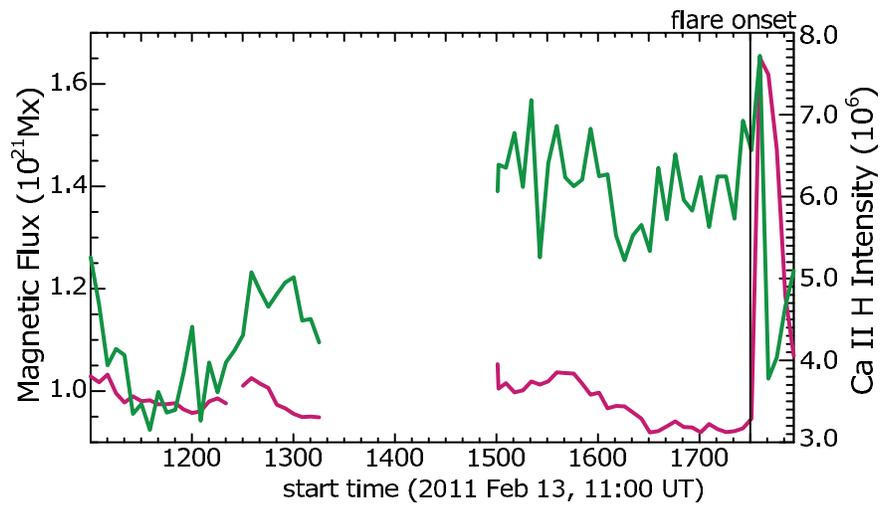}
\caption{
Temporal evolution of positive magnetic flux and Ca-line intensity prior to the M6.6 flare onset on NOAA AR 11158. 
The vertical solid line marks the onset time of the flare, 16:30 UT February 13, 2011. 
The image is formatted as described for Figure~\ref{fig:10930_flux}(b). 
The region of positive magnetic flux and Ca-line intensity is delineated by the yellow squares in Figures~\ref{fig:11158M6_dev}(a) and (e).
}
\label{fig:11158_flux}
\end{figure}

\begin{figure}
\plotone{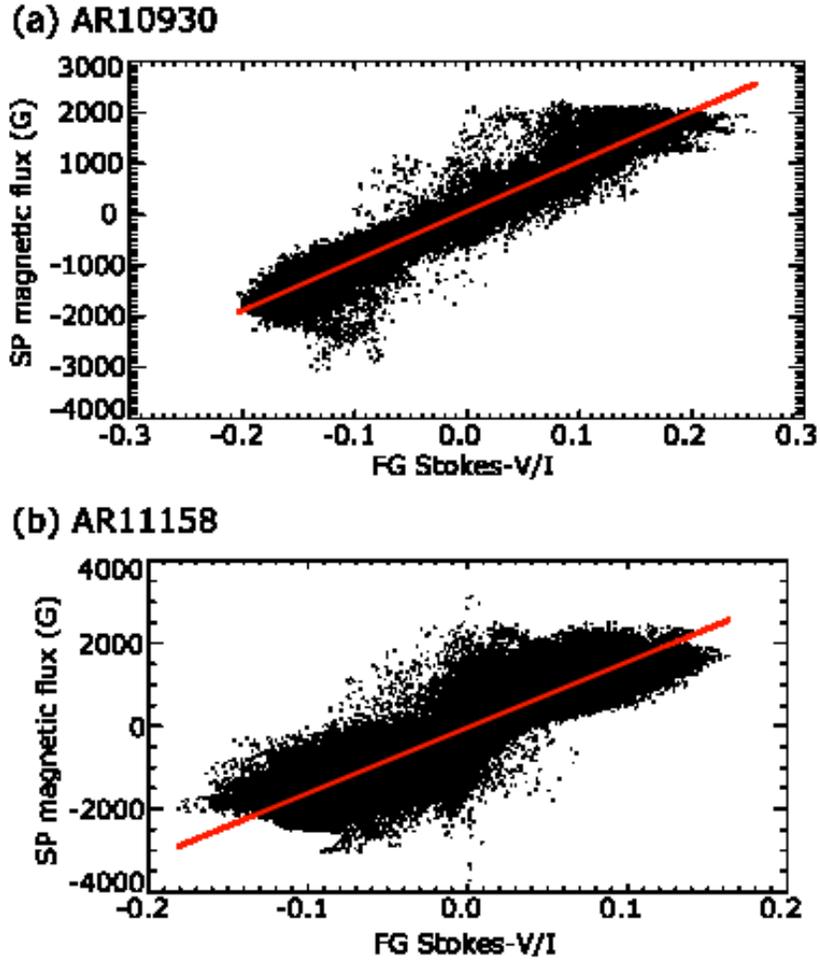}
\caption{
Intensity of LOS magnetic field by SP plotted against Stokes-V/I signal by FG for (a) X3.4 flare on NOAA AR 10930 and (b) M6.6 flare on NOAA AR 11158. 
The red lines were obtained by linear fitting. 
In (a), the LOS magnetogram scanned 04:30-05:36 UT and the filtergram at 04:12 UT December 13, 2006 were used. 
In (b), the LOS magnetogram scanned 16:00-16:32 UT, and filtergram obtained at 16:15 UT February 13, 2011 were used.
Intensity of Stokes-V/I signal was calibrated according to \citet{ichimoto08}.
}
\label{fig:sp_vs_fg}
\end{figure}

In order to investigate temporal evolution of positive magnetic flux on the flare-trigger region in Event 1, we converted the Stokes-V/I signal to magnetic field in Gauss units in terms of the LOS magnetogram scanned 04:00-05:36 UT on December 13, 2006, by SOT/SP and the filtergram at 04:44 UT December 13. 
Figure~\ref{fig:sp_vs_fg}(a) shows the scatter plots for them with the linear fitting line (red line).
We converted the Stokes-V/I signal to magnetic intensity $B$ (Gauss) by using this regression line, and calculated magnetic flux $F = \int B dS$ (Maxwell) plotted in Figure~\ref{fig:10930_flux}, where the integration is taken over the positive magnetic flux region in the yellow rectangle in Figure~\ref{fig:10930_flux}(a).

Magnetic flux of flare-trigger region in Event 3 was calculated by using LOS magnetogram scanned 16:00-16:32 UT on February 13, 2011, and the filtergram obtained at 16:15 UT February 13. 
The correlation map between the LOS magnetic field by SP and the Stokes-V/I signal by filtergram is shown with the linear fitting line in Figure~\ref{fig:sp_vs_fg}(b).


\clearpage

\begin{table}
\begin{center}
\caption{list of flare events}
\begin{tabular}{cccccc}
\tableline\tableline
event No. & date & start time\tablenotemark{a} & 
GOES X-ray class & active region & location\tablenotemark{b} \\
& & (UT) & & (AR NOAA) &\\
\tableline
1 & December 13, 2006 & 02:14 & X3.4 & 10930 & S07W22 \\
2 & December 14, 2006 & 22:07 & X1.5 & 10930 & S06W46 \\
3 & February 13, 2011 & 17:28 & M6.6 & 11158 & S20E05 \\
4 & February 15, 2011 & 02:44 & X2.2 & 11158 & S20W10 \\
\tableline
\label{table:eventlist}
\end{tabular}
\tablenotetext{a}{The start time is defined from X-ray observations of GOES satellite.}
\tablenotetext{b}{\url{http://www.solarmonitor.org/}}
\end{center}
\end{table}



\clearpage

\begin{deluxetable}{ccccc}
\tabletypesize{\scriptsize}
\rotate
\tablecaption{SOT/SP Data used in the analysis}
\tablewidth{0pt}
\tablehead{
\colhead{active region} & \colhead{date} & \colhead{SP scan time} & \colhead{field of view} & \colhead{puropose of use}\\
\colhead{(AR NOAA)} & & \colhead{(UT)} & &
}
\startdata
10930 & December 12, 2006 & 20:30:05 - 21:33:17  & 295$^{\prime\prime}$ $\times$ 162$^{\prime\prime}$ & measurement of the shear angle \\
10930 & December 13, 2006 & 04:30:05 - 05:36:08 & 295$^{\prime\prime}$ $\times$ 162$^{\prime\prime}$ & conversion of Stokes-V/I signels \\
10930 & December 14, 2006 & 17:00:05 - 18:03:17 & 295$^{\prime\prime}$ $\times$ 162$^{\prime\prime}$ & measurement of the shear angle \\
11158 & February 13, 2011 & 16:00:04 - 16:32:25 & 151$^{\prime\prime}$ $\times$ 162$^{\prime\prime}$ & measurement of the shear angle \\
 & & & & conversion of Stokes-V/I signels \\
11158 & February 14, 2011 & 06:30:04 - 07:02:25 & 151$^{\prime\prime}$ $\times$ 162$^{\prime\prime}$ & measurement of the shear angle \\
\enddata
\label{table:splist}
\end{deluxetable}

\end{document}